\documentclass[twocolumn,preprintnumbers,amsmath,amssymb]{revtex4-1}

\usepackage{graphicx}              
\usepackage{dcolumn}               
\usepackage{longtable}             
\usepackage{bm}                    
\usepackage{afterpage}

\def\beq{\begin{equation}}
\def\eeq{\end{equation}}

\begin{document}

\title{Basis set construction for molecular electronic structure theory: \\Natural orbital and Gauss-Slater basis for smooth pseudopotentials}

\author{F. R. Petruzielo$^{1}$}
\email{frp3@cornell.edu}
\author{Julien Toulouse$^{2}$}
\email{julien.toulouse@upmc.fr}
\author{C. J. Umrigar$^{1}$}
\email{CyrusUmrigar@cornell.edu}
\affiliation{
$^1$Laboratory of Atomic and Solid State Physics, Cornell University, Ithaca, New York 14853, USA\\
$^2$Laboratoire de Chimie Th\'eorique, Universit\'e Pierre et Marie Curie and CNRS, 75005 Paris, France\\
}

\date{\today}
\begin{abstract}
A simple yet general method for constructing basis sets for molecular electronic structure calculations is presented.
These basis sets consist of atomic natural orbitals from a multi-configurational self-consistent field calculation supplemented with primitive functions, chosen such that the asymptotics are appropriate for the potential of the system.
Primitives are optimized for the homonuclear diatomic molecule to produce a balanced basis set.
Two general features that facilitate this basis construction are demonstrated.
First, weak coupling exists between the optimal exponents of primitives with different angular momenta.
Second, the optimal primitive exponents for a chosen system depend weakly on the particular level of theory employed for optimization.
The explicit case considered here is a basis set appropriate for the Burkatzki-Filippi-Dolg pseudopotentials.
Since these pseudopotentials are finite at nuclei and have a Coulomb tail, the recently proposed Gauss-Slater functions are the appropriate primitives.
Double- and triple-zeta bases are developed for elements hydrogen through argon.
These new bases offer significant gains over the corresponding Burkatzki-Filippi-Dolg bases at various levels of theory.
Using a Gaussian expansion of the basis functions, these bases can be employed in any electronic structure method.
Quantum Monte Carlo provides an added benefit: expansions are unnecessary since the integrals are evaluated numerically. 
\end{abstract}
\maketitle
\section{Introduction}
\label{sec:intro}
In quantum chemistry (QC) calculations, molecular orbitals are traditionally expanded in a combination of primitive Gaussian basis functions and linear combinations of Gaussian primitives called contracted basis functions \cite{DunningJr1989}.
These basis sets cannot express the correct molecular orbital asymptotic behavior but are used in QC calculations to permit analytic evaluation of the two-electron integrals \cite{Boys1950}.

Analytic integral evaluation significantly limits flexibility in basis set choice but is essential for computational efficiency in QC calculations.
However, in practice, other basis function forms can be considered since an arbitrary function can be expanded in Gaussians.
Of course, the fidelity of this representation is limited.
An expansion in a finite number of Gaussians cannot reproduce the exponential decay of the wavefunction at large distances or the
Kato cusp conditions \cite{Kato1957} at nuclei, but it can mimic these features over a finite range.

Quantum Monte Carlo (QMC) calculations \cite{Foulkes2001} offer greater freedom in choice of basis functions because matrix elements are evaluated using Monte Carlo integration.
Consequently, the correct short- and long-distance asymptotics can be satisfied exactly.
For systems with a divergent nuclear potential, Slater basis functions can exactly reproduce the correct electron-nucleus cusp and long-range asymptotic behavior of the orbitals.
For calculations on systems with a potential that is finite at the nucleus and has a Coulomb tail, Gauss-Slater (GS) primitives \cite{Petruzielo2010} are the appropriate choice since they introduce no cusp at the origin and reproduce the exponential long-range asymptotic behavior of the orbitals.

Despite shortcomings, traditional QC basis sets have yielded good results.
The natural orbitals (NOs) from a post Hartree-Fock (HF) method are a particularly successful form of contracted function \cite{Almlof1987,Widmark1990,Widmark1991,Veryazov2004}.
The simplest NO construction involves diagonalizing the one-particle density matrix from a ground state atomic calculation \cite{Almlof1987}.
This construction is unbalanced due to obvious bias favoring the atom.
More complicated constructions involve diagonalizing the average one-particle density matrix of several systems: atomic ground and excited states, ions, diatomic molecules, and atoms in an external electric field \cite{Widmark1990,Widmark1991,Veryazov2004}.
These constructions produce excellent results, but they are complex.

A simple but general method for constructing basis sets for molecular electronic structure calculations is proposed and tested here.
The bases are combinations of the NOs obtained from diagonalizing the one-particle density matrix from an atomic multiconfigurational self-consistent field (MCSCF) calculation and primitive functions appropriate for the potential in the system.
The primitives are optimized for the homonuclear dimer in coupled cluster calculations with single and double excitations (CCSD),
with the intention of producing a balanced basis set.
Importantly, optimal exponents for the primitive functions are shown to depend weakly on the level of theory used in the optimization.
Additionally, results show that coupling is weak between primitive functions of different angular momenta.
This enables efficient determination of optimal exponents.

The utility of the above construction is demonstrated for the elements hydrogen through argon with the non-divergent pseudopotentials of Burkatzki, Filippi, and Dolg (BFD) \cite{Burkatzki2007}.
Since these pseudopotentials are finite at the nuclei and have a Coulomb tail, the GS functions are the appropriate primitives.
These pseudopotentials are chosen for demonstrated accuracy in all cases tested and because they are accompanied by a basis set.
The BFD basis \cite{Burkatzki2007} serves as a metric for testing the new basis.
The benefits of our bases extend to all electronic structure methods tested, including CCSD, HF, the Becke three-parameter hybrid density functional (B3LYP) \cite{Becke1993}, and QMC.

The main area of interest for the authors is QMC.
Since QMC results depend less on basis set than traditional QC methods \cite{Petruzielo2010}, only double-zeta ($2z$) and triple-zeta ($3z$) bases are presented.

This paper is organized as follows.
Basis function form and properties are demonstrated in Sec. \ref{sec:basis}.
Results for calculations with the new bases are discussed in Sec. \ref{sec:results}.
Concluding remarks are provided in Sec. \ref{sec:conc}.
Supplementary material is provided on EPAPS \cite{supplementary}.
\section{Basis Set}
\label{sec:basis}
The number of basis functions for each angular momentum follows the correlation consistent polarized basis set prescription of Dunning \cite{DunningJr1989}.
$2z$ and $3z$ bases appropriate for the BFD pseudopotentials are generated for the elements hydrogen through argon.
Since the BFD pseudopotential removes no core for hydrogen and helium, the $2z$ basis for these elements consists of two S functions and one P function, while the $3z$ basis consists of three S functions, two P functions, and one D function.
Since the BFD pseudopotential removes a helium core for the first row atoms and a neon core for the second row atoms, the remaining elements lithium through argon have the same number of basis functions.
In particular, the $2z$ basis consists of two S functions, two P functions, and one D function, while the $3z$ basis consists of three S functions, three P functions, two D functions, and one F function.

The bases consist of a combination of contracted and primitive functions.
Since the BFD pseudopotentials are finite at the origin and have a Coulomb tail, the GS functions are the appropriate primitives.
With the exception of the elements in Group $1A$ of the periodic table (i.e. H, Li, and Na), the basis for each element includes a single S contraction and a single P contraction combined with an appropriate number of GS primitives.
Only two contractions are employed to reduce the computational cost of using this basis in QC calculations.
Since elements in Group $1A$ of the periodic table have only one electron for the BFD pseudopotentials, a single S orbital is the ground state wavefunction, and this can be obtained exactly in HF.
Thus, the basis for each element in Group $1A$ includes a single S contraction, no P contractions, and an appropriate number of GS primitives.
\subsection{Contracted Functions}
\label{sec:contracted}
A contracted basis function is a linear combination of Gaussian primitives:
\beq
\varphi_{nlm}(r,\theta,\phi) = \sum_i c_i \;
\sqrt{\frac{2(2\alpha_i)^{n+\frac{1}{2}}}{\Gamma(n+\frac{1}{2})}} \;r^{n-1} e^{-\alpha_i r^2} Z_l^m(\theta,\phi),
\eeq
where $r,\theta,\phi$ are the standard spherical coordinates, $n$ is the principal quantum number, $l$ is the azimuthal quantum number,  $m$ is the magnetic quantum number, $Z_l^m(\theta,\phi)$ is a real spherical harmonic, $c_i$ is the $i^{\rm{th}}$ expansion coefficient, and $\alpha_i$ is the $i^{\rm{th}}$ Gaussian exponent.
In practice, the restriction $n=l+1$ applies.

The exponents of the primitive functions that form the contracted basis functions are determined as follows.
For each angular momentum for which a contraction is desired, an uncontracted basis consisting of nine even-tempered primitive Gaussians is generated.
For each set of uncontracted Gaussians, the minimum exponent and even-tempering coefficient are varied to minimize the CCSD energy of the atom using a Python wrapper around GAMESS \cite{Schmidt1993}.

An assumption of weak coupling between the different angular momenta underlies the optimization procedure.
Consequently, the uncontracted basis for each angular momentum is optimized separately.
This optimization is performed by calculating the CCSD energy on an initially coarse grid composed of different minimum exponents and even-tempering coefficients.
Once regions of low CCSD energy are identified, a finer grid is used to obtain the final minimum exponent and even-tempering coefficient.
In addition to the assumption of weak coupling, two other properties of the problem make this global optimization possible with modest computer resources; low dimensionality of search space and efficiency of atomic CCSD calculations.

Next, an atomic MCSCF calculation in a complete active space (CAS) with the optimized uncontracted basis is performed in GAMESS.
For these calculations, all electrons not removed by the pseudopotential are allowed to excite.
For helium, the active space consists of the orbitals from the $n=1$ and $n=2$ shells.
For beryllium through neon, the active space includes the orbitals from the $n=2$ and $n=3$ shells.
For magnesium through argon, the active space is composed of the orbitals from the $n=3$ and $n=4$ shells, with the exception of the $4$D and $4$F orbitals.
A subset of the natural orbitals from the MCSCF calculations are used as the contracted functions of our basis.

All atomic calculations are performed in $D_{2h}$ symmetry since GAMESS does not permit imposition of full rotational symmetry.
Hence, different components of the same atomic subshell are not necessarily equivalent.
Additionally, mixing may occur among orbitals of different angular momenta.
For instance, there is mixing of S orbitals with both D$_{3z^2-r^2}$ and D$_{x^2-y^2}$ orbitals.
This anisotropy can be removed by averaging the different components of a particular subshell
 and zeroing out the off-diagonal blocks of the one-particle density matrix \cite{Widmark1990}.

A simpler approach taken in this work is found to produce results of similar quality.
For each angular momentum for which a contraction is desired, the NO with that angular momentum which has the largest occupation number is chosen.
Additionally, NO elements which do not correspond to the dominant character of the orbital are zeroed out.
For instance, an NO with large coefficients on the S basis functions and small coefficients on the D basis functions is considered to be dominated by S character, so the D coefficients are zeroed out.
Finally, the NOs are normalized.
The NOs selected in this procedure generate the contracted functions for the basis set.
The expansions of the contractions are given in the supplementary material \cite{supplementary}.
\subsection{Gauss-Slater Primitives}
\label{sec:gs}
GS functions \cite{Petruzielo2010} are defined as
\beq
 \varphi^{\zeta}_{nlm}(r,\theta,\phi) =N_n^\zeta \; r^{n-1} e^{-\frac{(\zeta r)^2}{1+\zeta r}} \; Z_l^m(\theta,\phi),
\eeq
where $\zeta$ is the GS exponent and $N_n^\zeta$ is the normalization factor.
The restriction $n \ge l+1$ is imposed for GS functions.
For $r\ll 1$, the GS behaves like a Gaussian:
\beq
\varphi^{\zeta}_{nlm}(r,\theta,\phi) \cong N_n^\zeta \; r^{n-1} e^{-(\zeta r)^2} \; Z_l^m(\theta,\phi),
\eeq
and for $r\gg 1$, the GS behaves like a Slater:
\beq
\varphi^{\zeta}_{nlm}(r,\theta,\phi) \cong N_n^\zeta \; r^{n-1} e^{-\zeta r} \; Z_l^m(\theta,\phi).
\eeq
Consequently, GS functions introduce no cusp at the origin and can reproduce correct long-range asymptotic behavior of the orbitals.

Unlike Gaussians and Slaters, normalization of GSs has no closed form expression.
Nevertheless, normalizing an arbitrary GS is trivial with the following scaling relation between $N_n^\zeta$ and $N_n^1$:
\beq
N_n^\zeta = \zeta^{n+1/2} \; N_n^1.
\eeq
Values for $N_n^1$ are given in the supplementary material \cite{supplementary}.

Since GSs are not analytically integrable, the radial part must be expanded in Gaussians for use
in QC programs that evaluate matrix elements analytically.
The expansion is
\beq
\varphi^{\zeta}_{nlm}(r,\theta,\phi) = \sum_i c_i^\zeta \; 
\sqrt{\frac{2(2\alpha_i^\zeta)^{l+\frac{3}{2}}}{\Gamma(l+\frac{3}{2})}} \;r^l e^{-\alpha_i^\zeta r^2} Z_l^m(\theta,\phi),
\eeq
where $c_i^\zeta$ is the $i^{\rm{th}}$ expansion coefficient and $\alpha_i^\zeta$ is the $i^{\rm{th}}$ Gaussian exponent.
Notice that the expansion permits the case for which $n\neq l+1$ for the GS function.
Additionally, the following scaling relations hold for the expansion coefficients and Gaussian exponents:
\begin{align}
\alpha_i^\zeta &= \zeta^2 \alpha_i^1 \\
c_i^\zeta &= c_i^1.
\end{align}
Once the Gaussian expansions are found for unit exponents, expansions of arbitrary GSs follow immediately from the scaling relations.
For QC calculations in this paper, GSs are expanded in six Gaussians.
However, if the purpose of the initial QC calculation is to generate crude starting orbitals for QMC calculations in which orbital optimization is performed, it is only necessary to expand GS primitives in a single Gaussian.
In this case, the cost of QC calculations is the same for Gaussian and GS primitives.
The expansions of GS functions with unit exponent in both one and six Gaussians are given in the supplementary material \cite{supplementary}.

As mentioned above, the restriction $n \ge l+1$ is imposed for GS functions, instead of the more familiar $n=l+1$ restriction imposed for Gaussian primitives.
This motivates construction of two types of bases.
In the first, ANO-GS, the restriction $n=l+1$ is enforced.
In the second, ANO-GSn, for each $l$ there can be at most a single GS primitive with a particular $n$.
For each additional primitive with a particular $l$, $n$ must be incremented.

For example, consider lithium. 
The $2z$ ANO-GS basis has one S contraction, one GS-1S function, two GS-2P functions, and one GS-3D function.
On the other hand, the $2z$ ANO-GSn basis has one S contraction, one GS-1S function, one GS-2P function, one GS-3P function, and one GS-3D function.

A caveat to the above definition of the ANO-GSn basis is that GS-2S functions are not permitted since a single GS-2S function will introduce an undesired cusp in the wavefunction.
Additionally, the $2z$ ANO-GS and ANO-GSn basis sets are identical for all elements except lithium and sodium.
When the $2z$ ANO-GS and ANO-GSn basis sets are identical, the basis sets are referred to as a $2z$ ANO-GS/GSn basis.
For both lithium and sodium, the basis sets differ because these systems have no P contractions and instead have a second P primitive for the $2z$ basis.
This primitive is a GS-2P for the ANO-GS basis and a GS-3P for the ANO-GSn basis.
Additionally, weak coupling between functions of different angular momentum causes the GS-1S and GS-3D functions in the ANO-GS bases for lithium and sodium to differ from their counterparts in the ANO-GSn bases.
However, the optimal exponents differ by less than $0.01$.

Optimal exponent selection for the GS primitives is discussed now.
Instead of optimizing exponents for the atom as was done to generate the contractions, optimization of the GS exponents is performed for the homonuclear diatomic molecule at experimental bond length \cite{HubHer79, Gur79, NIST, CCCBDB, Grisenti2000, Bondybey1984, Aziz1989, Fu1990, Herman1988}. 
This advantageously produces a balanced basis set.

\begin{figure}[htp]
 \begin{center}
   \includegraphics[scale=0.70]{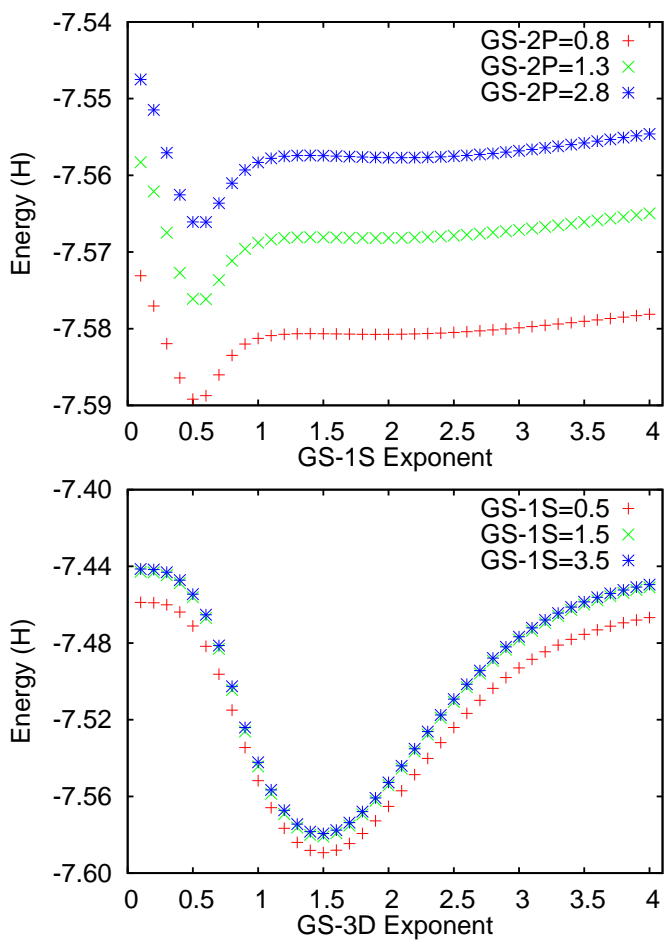}
   \caption{Change in Si$_2$ CCSD energy for $2z$ ANO-GS/GSn basis shows weak coupling between GS functions of different angular momenta.
     TOP: Energy versus GS-1S exponent for three values of the GS-2P exponent with the GS-3D exponent fixed at its optimal value.
     Bottom: Energy versus GS-3D exponent for three values of the GS-1S exponent with the GS-2P exponent fixed at its optimal value.}
   \label{fig:weak_coupling}
 \end{center}
\end{figure}

Weak coupling between GS functions of different angular momenta is assumed, so the initial optimization for each angular momentum is performed separately.
This assumption is validated in Figure \ref{fig:weak_coupling}, which contains plots of the CCSD energy for Si$_2$ while varying individual GS exponents in the $2z$ ANO-GS/GSn basis.
Both the curve shape and exponent value which minimizes the energy vary little with fixed exponent value, signifying weak coupling between GS functions of different angular momentum.

\begin{figure}[htp]
 \begin{center}
   \includegraphics[scale=0.70]{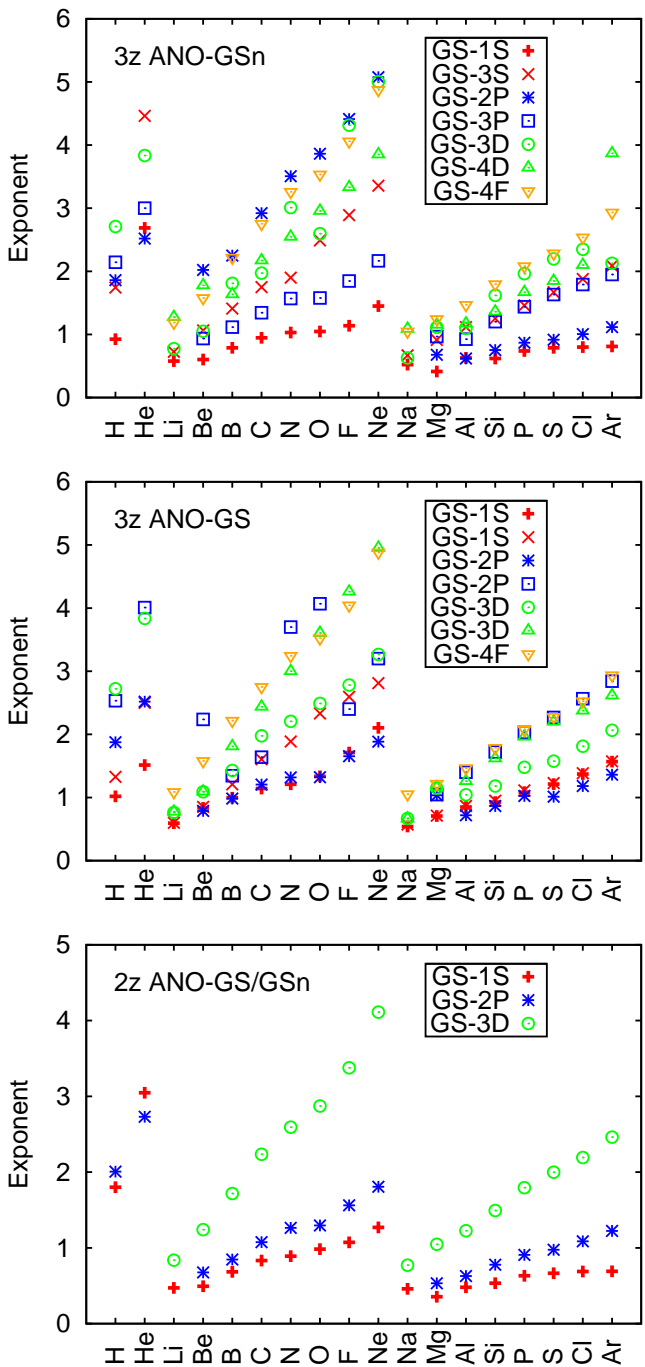}
   \caption{Optimal exponents for ANO-GS and ANO-GSn bases exhibit a linear trend across each row of the periodic table.
     The $2z$ ANO-GS and ANO-GSn bases are identical for all elements except lithium and sodium.
     The GS-1S and GS-3D exponents for these elements each differ by less than $0.01$ between $2z$ ANO-GS and ANO-GSn bases, so $2z$ ANO-GS and ANO-GSn are shown together as $2z$ ANO-GS/GSn.
     Exponents for GS functions of P angular momentum are not included for lithium and sodium since these elements have an extra primitive of P angular momentum.}
   \label{fig:exponents}
 \end{center}
\end{figure}

The optimization is performed at the CCSD level of theory using a Python wrapper around GAMESS.
For each angular momentum, an energy landscape is defined by a grid of primitive exponents ranging from $0.1$ to $6.0$ with $0.1$ spacing.
Thorough investigation has revealed that exponents larger than $6.0$ are not optimal for the systems considered.
Low lying minima of this energy landscape are then handled with increasingly finer grids until energy changes are less than $0.01$ mH.
During this investigation of local minima, all angular momenta are handled simultaneously to account for any coupling effects.
Results of this optimization are shown in Figure \ref{fig:exponents}.
Optimal exponents for ANO-GS and ANO-GSn bases exhibit a linear trend across each row of the periodic table.
For nearly degenerate minima, the exponent following the trend in the figure is chosen as optimal, resulting in energy increase no greater than several $0.1$ mH.
The optimal GS exponents are given in the supplementary material \cite{supplementary}.

In some cases, the optimal exponents for primitives with the same $n$ and $l$ are very close.
This can lead to large equal and opposite coefficients on these basis functions when constructing molecular orbitals.
Numerical problems could result, providing further motivation for the ANO-GSn basis, in which each pair of $n$ and $l$ is unique.
However, all of our tests with the ANO-GS basis have had no numerical problems.

\begin{figure}[htp]
 \begin{center}
   \includegraphics[scale=0.70]{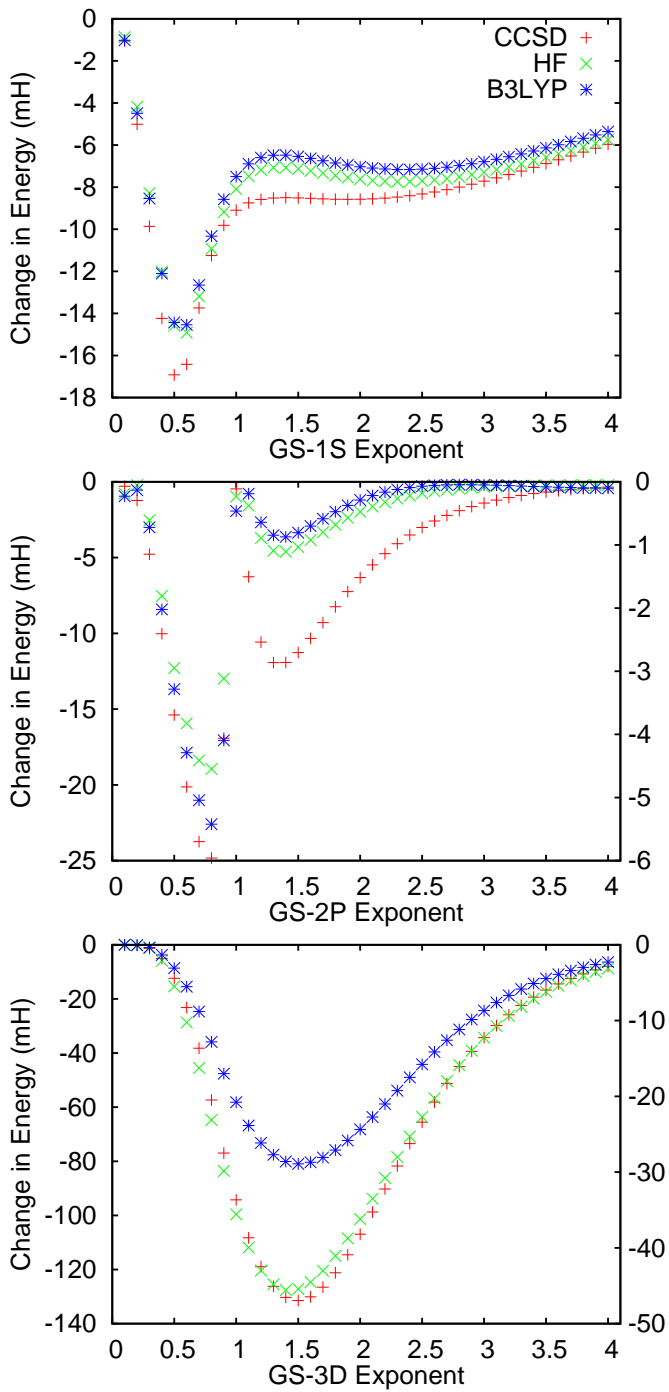}
   \caption{Change in Si$_2$ energy for $2z$ ANO-GS/GSn basis shows optimal exponents depend weakly on electronic structure method (CCSD, HF, and B3LYP).
     Top: GS-1S exponent is varied with GS-2P and GS-3D exponents fixed at their optimal values.
     Middle: GS-2P exponent is varied with GS-1S and GS-3D exponents fixed at their optimal values. The large increase in energy around an exponent of $1.0$ occurs since the P primitive and P contraction become nearly linearly dependent. 
     Bottom: GS-3D exponent is varied with GS-1S and GS-2P exponents fixed at their optimal values.
     For Middle and Bottom, HF and B3LYP energy scale is on the right y-axis.
     This difference in energy scale occurs since higher angular momentum functions are less important in these effectively single-determinant theories.}
   \label{fig:testing_different_theories}
 \end{center}
\end{figure}

Finally, the optimal primitive exponents are found to depend weakly on the electronic structure method employed in the
optimization, as demonstrated in Figure \ref{fig:testing_different_theories} for Si$_2$ with the $2z$ ANO-GS/GSn basis.
The globally minimizing exponents are nearly equal in different methods.
This exponent transferability to different levels of theory is extremely attractive for a basis set.

\section{Results}
\label{sec:results}
Section \ref{sec:basis} demonstrates that the ANO-GS and ANO-GSn bases exhibit desirable properties.
However, it remains to be shown that these basis sets produce accurate results.
Fortunately, the basis set accompanying the BFD pseudopotential serves as a metric for testing ANO-GS and ANO-GSn basis quality.
The BFD basis for elements in Groups $1A$ and $2A$ of the periodic table has recently been updated \cite{BFD}, 
but the number of functions in the new basis is inconsistent with the correlation consistent polarized basis prescription \cite{DunningJr1989}.
Since comparison would be difficult, their published functions are considered in this work.

\begin{figure}[htp]
 \begin{center}
   \includegraphics[scale=0.70]{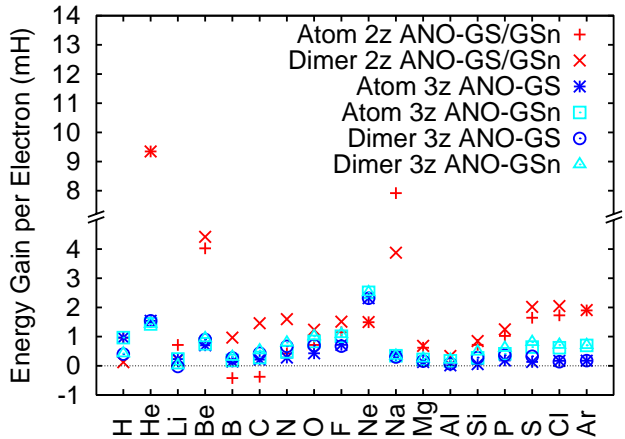}
   \caption{CCSD total energy gains per electron of ANO-GS and ANO-GSn relative to the corresponding BFD basis \cite{Burkatzki2007} for atoms and homonuclear dimers of hydrogen through argon.
     Energy gains per electron tend to increase across each row of the periodic table.
     The $2z$ ANO-GS and ANO-GSn bases are identical for all elements except lithium and sodium.
     Differences between $2z$ ANO-GS and ANO-GSn results for these elements is $\sim 0.01 $mH, so they are shown together as $2z$ ANO-GS/GSn.}
   \label{fig:energy}
 \end{center}
\end{figure}

Figure \ref{fig:energy} shows the CCSD total energy gain per electron of the ANO-GS and ANO-GSn bases over the BFD bases \cite{Burkatzki2007} for atoms and homonuclear dimers of hydrogen through argon.
Energy gains per electron tend to increase across each row of the periodic table.
Both ANO-GS and ANO-GSn bases yield energy gains for most molecules and atoms.
The energy gains per electron are generally larger for molecules than for atoms, and larger for the ANO-GSn basis than for the ANO-GS basis.
The energy gains for the $2z$ bases are generally larger than for the $3z$ bases, as expected, since the energy left to recover becomes smaller as the basis size increases.

\begin{figure*}[htp]
 \begin{center}
   \includegraphics[scale=0.70]{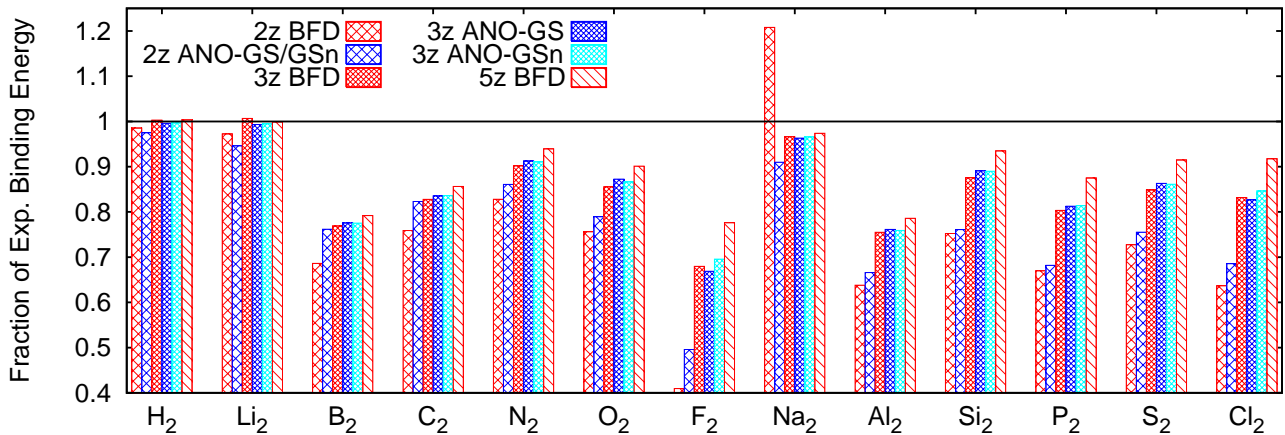}
   \caption{Fraction of experimental atomization energy recovered in CCSD with BFD, ANO-GS, and ANO-GSn bases for the homonuclear dimers of hydrogen through argon which are not weakly bound.
     The $2z$ ANO-GS and ANO-GSn bases are identical for all elements except lithium and sodium.
     Differences between $2z$ ANO-GS and ANO-GSn atomization energies for these elements is $\sim 0.01 $mH, so they are shown together as $2z$ ANO-GS/GSn.
     Calculated values are corrected for zero point energy \cite{Irikura2007, CCCBDB} to compare with experiment \cite{CCCBDB, HubHer79, NIST, Luo2007}.}
   \label{fig:fraction_recovered_bar_horizontal}
 \end{center}
\end{figure*}

The ANO-GS and ANO-GSn bases also produce more accurate CCSD atomization energies than the BFD basis for the homonuclear dimers of hydrogen through argon.
Figure  \ref{fig:fraction_recovered_bar_horizontal} shows the fraction of experimental atomization energy recovered in CCSD for the homonuclear dimers which are not weakly bound.
The $2z$ ANO-GS/ANO-GSn basis recovers more atomization energy than the $2z$ BFD basis for all dimers except those of Group $1A$ elements.
Similarly, the $3z$ ANO-GSn basis recovers more atomization energy than the $3z$ BFD basis for the same systems, but the differences are small.
The $3z$ ANO-GSn is on average slightly better than the $3z$ ANO-GS basis, the largest gains being for F$_2$ and Cl$_2$.

For Group $1A$ elements, the BFD bases recover more atomization energy in CCSD than do their ANO-GS or ANO-GSn counterparts.
This occurs due to inaccurate BFD energies for the atoms, as can be seen in Figure \ref{fig:energy}.
However, as described above, we used the published BFD bases for these elements rather than the updated BFD bases \cite{BFD} to maintain consistency.

\begin{figure}[htp]
 \begin{center}
   \includegraphics[scale=0.70]{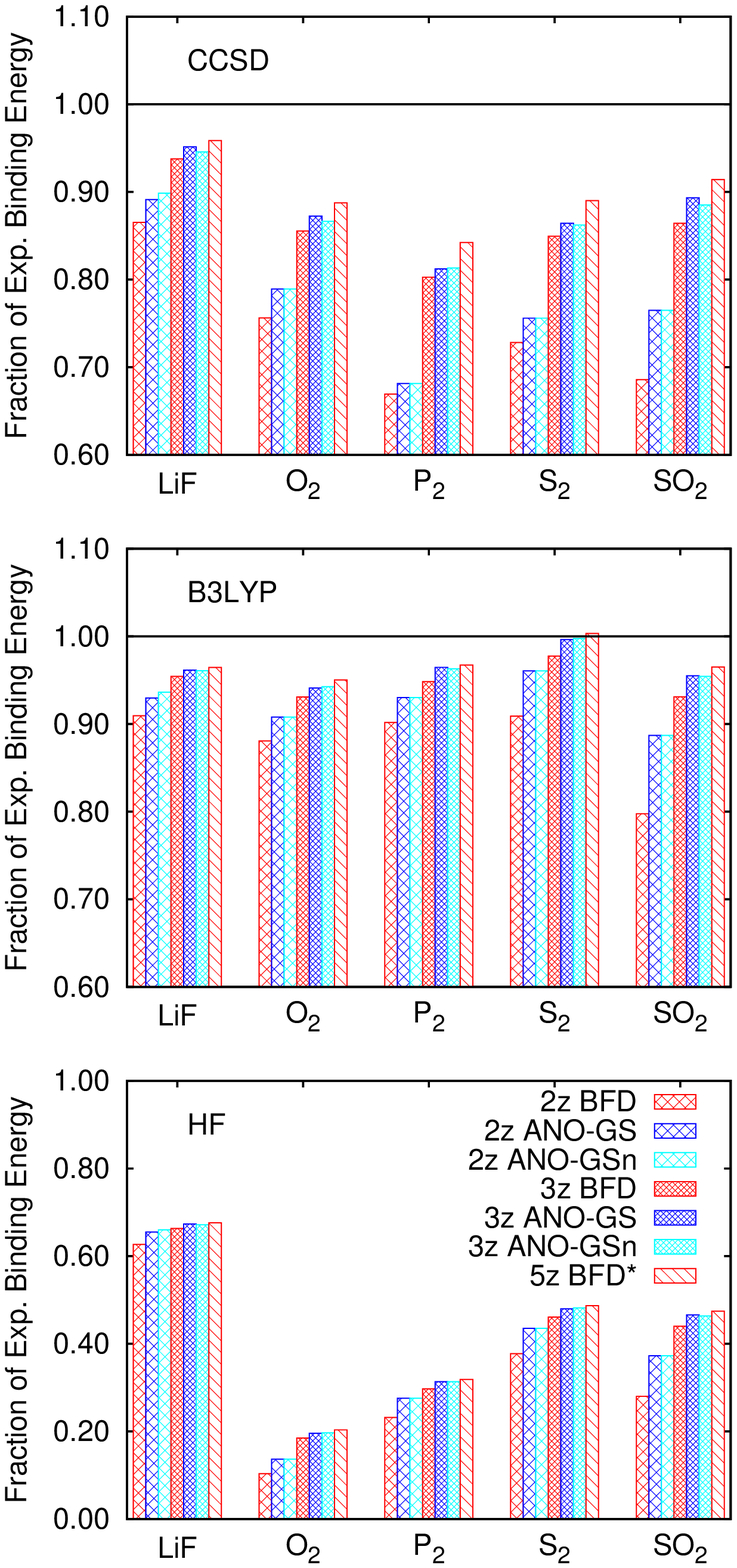}
   \caption{
  Fraction of experimental atomization energy recovered in HF, B3LYP, and CCSD for LiF, O$_2$, P$_2$, S$_2$, and SO$_2$ with BFD, ANO-GS, and ANO-GSn bases.
  The $2z$ ANO-GS and ANO-GSn bases yield different results only for LiF.
  The $5z$ BFD* calculations do not include the G or H functions from the $5z$ BFD basis.
  Calculated atomization energies are corrected for zero point energy \cite{Irikura2007, CCCBDB} to compare with experiment \cite{CCCBDB, Feller1999, HubHer79, NIST}.}
   \label{fig:g2_binding_qc}
 \end{center}
\end{figure}

Finally, improvements of the ANO-GS and ANO-GSn bases extend to other systems and methods.
Figure \ref{fig:g2_binding_qc} shows the fraction of experimental atomization energy recovered for five systems in the G2 set \cite{Curtiss1991} with the BFD, ANO-GS, and ANO-GSn bases in three quantum chemistry methods.
For CCSD, the ANO-GS and ANO-GSn bases outperform the BFD basis for all systems.
For sulfur dioxide the improvement due to the ANO-GS and ANO-GSn bases is dramatic: the $2z$ ANO-GS/GSn result is nearly halfway between the $2z$ and $3z$ BFD results, and the $3z$ ANO-GS/GSn result is nearly halfway between the $3z$ and $5z$ BFD results.
ANO-GS and ANO-GSn benefits are more prominent in HF and B3LYP: for most systems, the $2z$ ANO-GS/GSn result is closer to the $3z$ BFD result than the $2z$ BFD result,  and the $3z$ ANO-GS/GSn result is closer to the $5z$ BFD result than the $3z$ BFD result.
Differences between results with the ANO-GS and ANO-GSn bases are small.

\begin{figure}[htp]
 \begin{center}
   \includegraphics[scale=0.70]{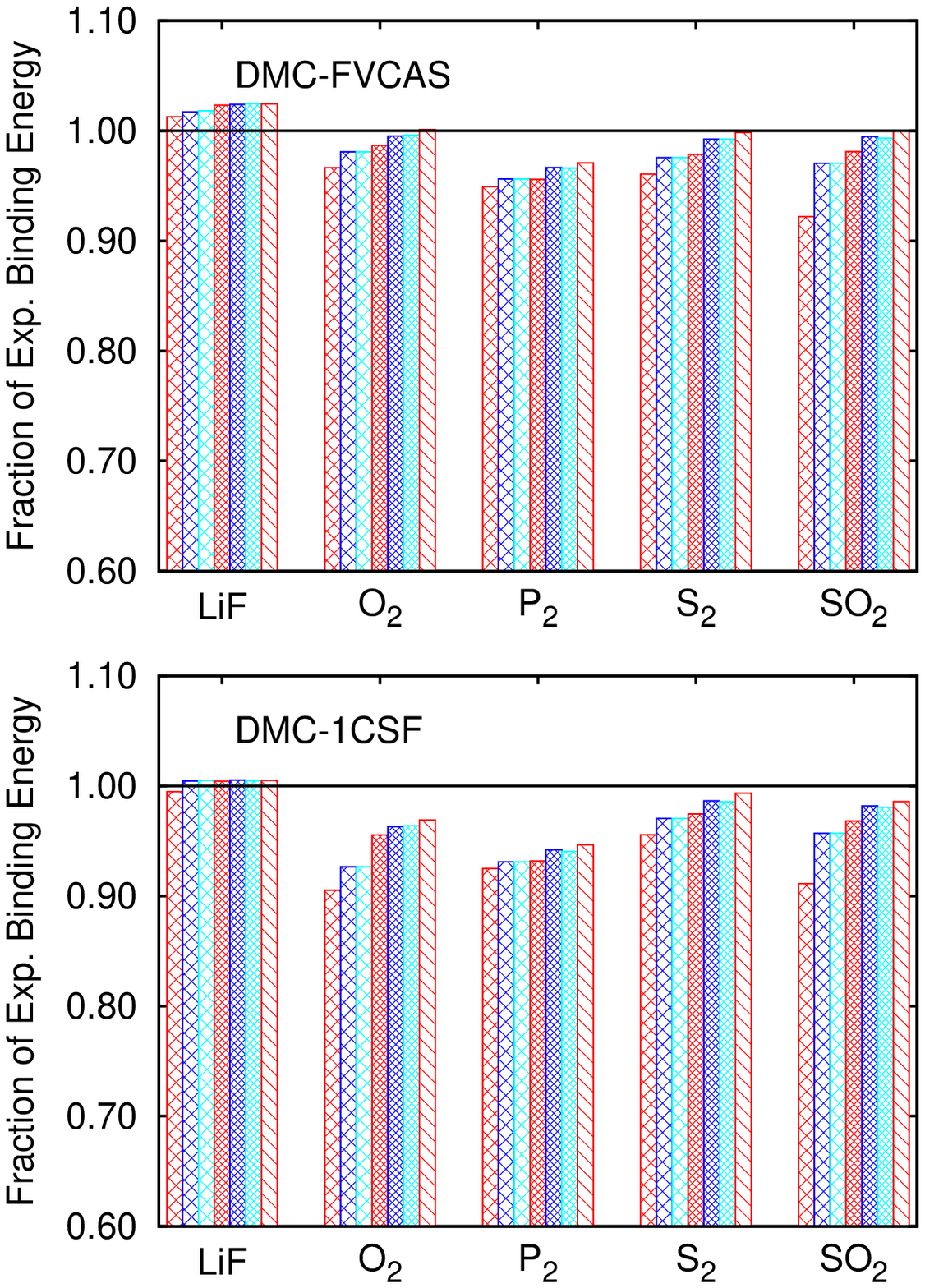}
\caption{
  Fraction of experimental atomization energy recovered in diffusion Monte Carlo (DMC) for LiF, O$_2$, P$_2$, S$_2$, and SO$_2$ with the BFD, ANO-GS, and ANO-GSn bases.
  DMC calculations are performed with both a single-CSF reference (DMC-1CSF) and full-valence complete active space reference (DMC-FVCAS).  
  The $2z$ ANO-GS and ANO-GSn bases yield different results only for LiF.
  The $5z$ BFD* calculations do not include the G or H functions from the $5z$ BFD basis.
  Calculated atomization energies are corrected for zero point energy \cite{Irikura2007, CCCBDB} to compare with experiment \cite{CCCBDB, Feller1999, HubHer79, NIST}.
  The legend for this plot is identical to that of Figure \ref{fig:g2_binding_qc}.}
   \label{fig:g2_binding_qmc}
 \end{center}
\end{figure}

Figure \ref{fig:g2_binding_qmc} shows the fraction of experimental atomization energy recovered using diffusion Monte Carlo (DMC) with the BFD, ANO-GS, and ANO-GSn bases.
For each system, the DMC calculations are performed with both a single-configuration state function (single-CSF) reference (DMC-1CSF) and full-valence complete active space reference (DMC-FVCAS).
However, for each of the constituent atoms in these molecules, the FVCAS and single-CSF references are equivalent.
All DMC calculations are performed with a $0.01$ H$^{-1}$ time step and trial wavefunction obtained by optimizing Jastrow, orbital, and configuration state function (CSF) parameters (where applicable) via the linear method \cite{Toulouse2007,Toulouse2008,Umrigar2007} in variational Monte Carlo.  
The DMC-1CSF and DMC-FVCAS calculations exhibit similar trends to the HF and B3LYP calculation for most systems: the $2z$ ANO-GS/GSn result is closer to the $3z$ BFD result than the $2z$ BFD result,  and the $3z$ ANO-GS/GSn result is closer to the $5z$ BFD result than the $3z$ BFD result.
Again, differences between results with the ANO-GS and ANO-GSn bases are small.

There are several important points that can be made by comparing the DMC calculations of Figure \ref{fig:g2_binding_qmc} 
to the CCSD calculations of Figure \ref{fig:g2_binding_qc}.
First, the DMC results for the atomization energies have a weaker dependence on basis size than the CCSD results.
Second, for a given basis set, the most basic DMC calculations, DMC-1CSF, yield superior results compared to CCSD.
In addition to yielding superior results, DMC-1CSF calculations have better computational cost scaling than CCSD calculations.
Under certain assumptions, the cost of DMC-1CSF calculations scales as $\mathcal{O}(N^3)$ \cite{Needs2010},  
while the cost of CCSD calculations scales as $\mathcal{O}(N^6)$ \cite{Helgaker2000}, where $N$ is the number of electrons.
However, it is important to note that the prefactor of the scaling is significantly smaller for the CCSD calculations.

Finally, our results are not the first to show that DMC calculations can produce accurate atomization energies.
In particular, DMC-1CSF calculations of the entire G2 set have been performed for both pseudopotential and all-electron systems \cite{Grossman2002,Nemec2010} and produced excellent results.
Additionally, there is good agreement between the pseudopotential and all-electron results with a mean absolute deviation of about $2.0$ kcal/mol over the entire G2 set \cite{Nemec2010}.
Although these previous results are very good, there is room for improvement, particularly for the open shell systems.
A systematic study with DMC-FVCAS calculations is currently underway in our group, which should produce results to (near) chemical accuracy for all systems in the G2 set.
\section{Conclusion}
\label{sec:conc}
A simple yet general method for constructing basis sets for molecular electronic structure theory calculations has been presented.
These basis sets consist of a combination of atomic natural orbitals from an MCSCF calculation with primitive functions optimized for the corresponding homonuclear dimer.
The functional form of the primitive functions is chosen to have the correct asymptotics for the nuclear potential of the system.

It was shown that optimal exponents of primitives with different angular momenta are weakly coupled.
This enables efficient determination of optimal exponents.
Additionally, it was demonstrated that the particular electronic structure method employed in optimization has little effect on the optimal values of the primitive exponents.

Two sets of $2z$ and $3z$ bases, ANO-GS and ANO-GSn, appropriate for the Burkatzki, Filippi, and Dolg non-divergent pseudopotentials were constructed for elements hydrogen through argon.
Since these pseudopotentials do not diverge at nuclei and have a Coulomb tail, GS functions are the appropriate primitives.

It was demonstrated that both ANO-GS and ANO-GSn basis sets offer significant gains over the Burkatzki, Filippi and Dolg basis sets for CCSD, HF, B3LYP \cite{Becke1993}, and QMC calculations.
Improvements were observed in both total energies and atomization energies.
The latter indicates that basis sets providing a balanced description of atoms and molecules were produced by using both the atom and the dimer in the optimization.
On average, the ANO-GSn basis is slightly better than the ANO-GS basis, but either is a sound choice.

In the future, these basis sets will be extended to include the transition metals, and,
bases will be constructed for all-electron calculations, for which Slater functions are the appropriate primitives.
\section{Acknowledgments}
We thank Claudia Filippi for very valuable discussions.
This work was supported by the NSF (Grant Nos. DMR-0908653 and CHE-1004603).
Computations were performed in part at the Computation Center for Nanotechnology Innovation at Rensselaer Polytechnic Institute.
\clearpage
\bibliographystyle{apsrev4-1}
\bibliography{contractions_gs}
\end{document}